\documentstyle[aps,prl,preprint]{revtex}

\begin{document}

\draft

\title{ Dislocation Emission around Nanoindentations on
a (001) fcc Metal Surface Studied by STM and Atomistic Simulations}

\author{$^1$O. Rodr\'{\i}guez de la Fuente\cite{corresp}, 
$^2$J. A. Zimmerman, $^1$M. A. Gonz\'{a}lez, 
$^2$J. de la Figuera, $^2$J. C. Hamilton, $^3$Woei Wu Pai, $^1$J. M. Rojo}

\address{
$^1$Departamento de F\'{\i}sica de Materiales, Universidad Complutense, Madrid 28040, Spain\\
$^2$Sandia National Laboratories, Livermore, California 94550, USA\\
$^3$Center for Condensed Matter Sciences, National Taiwan University, Taiwan 106
}

\date{\today}

\maketitle

\begin{abstract}
We present a combined study by Scanning Tunneling Microscopy and 
atomistic simulations of the emission of dissociated dislocation 
loops by nanoindentation on a (001) fcc surface. The latter consist 
of two stacking-fault ribbons bounded by Shockley partials and a 
stair-rod dislocation. These dissociated loops, which intersect the 
surface, are shown to originate from loops of interstitial character
emitted along the \textless110\textgreater\ directions and are 
usually located at hundreds of angstroms away from the indentation 
point. Simulations reproduce the nucleation and glide of these 
dislocation loops.
\end{abstract}


Nanoindentation is a well-recognized
technique that elucidates mechanical properties of surfaces and thin
films at atomic scale \cite{nanoindentation_review,nanoAu2,nanoAu}. Although 
it is the small-scale counterpart of the traditional hardness test, it is
far from being understood at the same level. Experimental evidence is
still rather fragmentary although a number of studies involving
atomic force microscopy and related techniques
have provided information about such remarkable phenomena as the 
passivating properties of self-assembled molecular single layers
\cite{passivating} or the softening role of steps in the onset 
of plasticity \cite{kiely98}. Although most of
our initial knowledge of the processes taking place under a
nanoindenter comes from molecular dynamics simulations, in which
much progress has been made in recent years \cite{cindy98,tadmor99},
a limiting factor of the comparison of atomistic simulations
and experiment has been the different length scales accessible to each.
In the present work we have succeeded to bridge the length-scale
gap by increasing the experimental resolution around the 
nanoindentation point down to atomic scale.

The nucleation of discrete dislocations and their interactions are
crucial to understand the first steps of crystal plasticity.
In the present Letter, we report the first direct observation by
Scanning Tunneling Microscopy (STM) of the emission of 
individual dislocation
loops by nanoindentation of a Au(001) surface and show
that a combined STM and atomistic simulation investigation 
explains how these loops are originated and how they glide 
away hundreds of angstroms from the indentation point. 
We fully characterize these loops and present evidence of 
the atomic displacement processes that are responsible for their 
generation. Furthermore, standard 
dislocation theory is shown to provide a suitable and convenient 
framework for encompassing the experimental results and the atomic 
simulations.

The experiments were performed in ultra high vacuum conditions on a
Au(001) crystal cleaned by repeated sputtering and annealing in a
system described in more detail elsewhere
\cite{au001_dislocaciones}. Nanoindentations were done pushing the 
tungsten tip against the sample for distances of a few nanometers
from the stable tunnel distance with the control feedback switched off.
The last layer of the Au(001) surface is reconstructed with the well known 
``hex'' $5\times20$ reconstruction which appears in STM 
images as fringes oriented along a \textless110\textgreater\ direction. 
These fringes arise from a moir{\'e}-like effect between the topmost layer 
with an hexagonal arrangement and the lower bulk-like layer with square 
symmetry.

A typical image of the surface of the crystal after performing
nanoindentations is shown in Fig.~\ref{fig1}.
Indentations themselves are imaged as multi-storied pits, and some 
material is shown as a pile-up surrounding the pits. Rows of bump-like 
features with a height of 0.6 $\pm0.1$ {\AA} (called from now on {\it hillocks}) 
are apparent along the compact \textless110\textgreater\ 
directions on the (001) surface, at distances of hundreds of
angstroms from the indentation points.
We have also observed similar {\it hillocks} in the course of ion
irradiation on both Au(001) and
Ag(001) followed by gentle annealing: in these cases the
{\it hillock} distribution is random, with no alignments in rows as the ones
surrounding nanoindentations.

Close-up STM images of {\it hillocks} are shown in Fig.~\ref{fig2}
together with an interpretation of their sub-surface
structure\cite{au001_dislocaciones}, which consists of two
stacking-faults on intersecting \{111\} planes
each bounded by two parallel Shockley partial dislocations.
The whole configuration is held up by a stair-rod dislocation
parallel to the surface. We argue that
the origin of the {\it hillocks} can be traced to perfect dislocation 
loops that are punched into the crystal by the tip displacement into 
the surface \cite{oscarMRS00}. In our case, a family of such 
dislocation loops would consist of V-shaped half-loops intersecting
the surface with Burgers vector parallel to the latter. They would, on
energetic grounds, be split into two pairs of Shockley partials giving
rise to the configuration shown in Fig.2b, i.e. resulting in
{\it hillocks}\cite{thompson}. 
We can ascertain that these loops
are of interstitial character based on the fact that for Au (Fig. 2a) 
we observe a missing reconstruction fringe on top of the {\it hillock}. 
As the interatomic spacing of the substrate is larger 
than the one of the uppermost reconstructed layer, the position of the 
missing-fringe in the moir\'e-like pattern\cite{hirsch} corresponds to 
the position of the extra row of interstitial atoms below.
Furthermore, for the Ag(001) atomic resolution image of Fig.~\ref{fig2}c, 
it is clear that each partial dislocation produces a mismatch of one half 
interatomic unit between the rows on both sides of each stacking-fault, 
the two mismatches adding up to one extra row of atoms in the 
inner side of the {\it hillock}.

To gain insight into the atomic processes involved in the creation of
these {\it hillocks}, atomistic simulations\cite{details} were carried
out. A repulsive potential\cite{cindy98} was used to model a spherical 
indenter of radius 4 nm penetrating the surface of a Au(001) crystal
modeled with the embedded atom method potential
\cite{EAM}. The reconstructed layer is thought to behave like a 
floating layer \cite{fiorentini93}. Thus, it  
is not expected to affect dislocation generation and behavior during
nanoindentation and was not included in the simulations.
A top view of the surface after the simulated 
nanoindentation is shown in Fig.~\ref{fig3}. In agreement with
experiment, it is observed that two
{\it hillocks} have been generated around the nanoindentation trace.
To make a more quantitative comparison between simulations and 
dislocation models, a quantity closely related to the Burgers vector 
can be defined for every atom\cite{zimmerman00}. 
This quantity is called the slip-vector, ${\bf s}_i$,
being defined for atom $i$ as:
\begin{equation}
        {\bf s}_i = - \frac{1}{N_s}\sum^{N_{nn}}_{j\neq i}({\bf r}_{ij}-{\bf r}^0_{ij})
\end{equation}
where ${\bf r}_{ij}$ and ${\bf r}^0_{ij}$ are the vectors linking atom $i$ and all
its $N_{nn}$ nearest neighbors $j$ in their current and
reference (prior to the indentation) positions, respectively. 
$N_s$ stands for the number of
slipped neighbors. The spatial distribution of the slip vectors moduli
$\mid{\bf s}_i\mid$ around the nanoindentation trace, with a suitable
color scale, is shown in Fig.~\ref{fig4}. Green atoms with
$\mid{\bf s}_i\mid \sim \frac{a_0}{\sqrt{6}}$ are atoms on a stacking
fault ($a_0$ is the Au lattice parameter). Blue atoms correspond 
to values of the slip-vector modulus between the 
stacking fault value and zero, while yellow ones have a slip-vector
ranging between a complete lattice parameter and the stacking fault value.
Blue and yellow atoms are, consequently, 
around the core of a Shockley partial dislocation. The configuration
shown in Fig.~\ref{fig4}, 
with a height displacement for the topmost atoms of 0.6 {\AA}, strikingly 
reproduces the previously proposed subsurface structure of a {\it hillock}.

A dynamical picture of the process is obtained by
recording successive frames of the simulated atomic events.
To obtain the sequence, the
indenter is first lowered in increments of  0.01 nm down to a depth of
0.58 nm in a quasi-static way at zero temperature. At this point
dislocation loops are observed below the indenter, in agreement
with previous results\cite{cindy98}. Then, a constant energy molecular 
dynamics simulation follows the evolution of the system for 36 ps. A 
different dislocation configuration (again reproducing the previously 
proposed subsurface structure of a {\it hillock}) is created close to 
the tip and, then, glides away in a \textless110\textgreater\ direction.
We stress that, although the experimentally observed {\it hillocks} 
are usually much larger than the simulated ones (and the indentation
itself is also much deeper), we observe also {\it hillocks} of the 
same size in both experiment and simulation.

{\it Hillocks} are seen to glide as a whole unit.
This behavior can be understood on the grounds
that Shockley partial dislocations bounding a stacking fault are
expected to glide easily on the \{111\} gliding planes. In our
simulation they indeed glide away dragging with them the stair-rod
dislocation (motion of structures formed by stair-rod
dislocations and Shockley partial dislocations in thin films has been
recently reported\cite{glide01}). We argue that the rows of {\it hillocks}
appearing in the STM image of Fig.~\ref{fig1} are the
result of successive emission of loops that glide away from the nanoindentation
trace. Once started into motion due to the high stress close to the
indentation, the {\it hillocks} would glide away from the indentation point
until they collide or interact with other defects in the crystal. The
simulation cell is too small to observe in detail this effect, although
the {\it hillocks} move with a constant velocity within the unit cell once
they are far enough from the indentation point. {\it Hillock}-like
structures (although interpreted in a different way) have
been observed to drift in highly stressed regions\cite{gai96}.
The creation of
{\it hillocks} following ion irradiation can also be explained in terms of
the above model: after long ion irradiation and further annealing,
the surface is known to
exhibit a multi-storied pit structure\cite{orf}, whereas the presence
of a large supersaturation of adatoms and, probably, sub-surface
interstitials, is suspected. These extra-atoms can cluster on crystallographic
planes and, after relaxation, give rise to perfect loops that start
the above mechanism.

In the course of the many simulations performed, a variety of {\it hillock}
structures have been found. They differ in the exact arrangement of
the loop below the surface: more complex configurations than the
simplest one observed in Fig.~\ref{fig4} (an edge V-shaped loop
dissociated along \{111\} planes) are possible. But the defining
characteristics are the same for all of them: the Burgers vector of
the undissociated loop is a lattice vector, all the sections of the loop
appear dissociated along \{111\} planes, and the loop glides as a whole unit
in a \textless110\textgreater\ direction. It is also worth remarking that, 
in the simulations, the {\it hillocks} remain in place once the tip is 
retracted from the surface, in agreement with the experiment.

Comparing different experimental images, it is observed that
the size distribution of both the span $s$ (distance between stacking
faults at the surface), and width $w$ (separation between partial
dislocations again at the surface) of the observed {\it hillocks} is rather
broad, ranging from a lowest resolvable size of about one
reconstruction period ($\sim$ 1.4 nm) to a size of several
ones. However, the parameter $w$ (in fact, the width of the extended
dislocation) is experimentally seen to level off with increasing
span $s$ of the {\it hillock}. The repulsive interaction force 
between two Shockley partial dislocation 
segments, at a given distance, increases initially with segment
length (which is proportional to the observed span)
but attains a constant value when this length becomes much
larger than the distance between the segments. Within a model using
elemental dislocation theory\cite{hirth} and taking into account image 
dislocations to include the effect of the free surface, the exact form 
of the $w(s)$ curve can be predicted and is found to be in agreement 
with the experimental data\cite{unpub}. It is worth emphasizing that the 
parameters (Burgers vectors, geometry...) of this new configuration 
can be explained in terms of continuum dislocation theory.

In summary, we have shown that the initial stages of plastic deformation
around a nanoindentation result in the emission from near the contact 
point of dislocation half-loops intersecting the surface; they can, 
alternatively, be created by accretion of irradiated-in interstitials. 
These loops are split into pairs of Shockley partial dislocations
giving rise to peculiar configurations at the surface ({\it hillocks}),
involving four Shockley partial dislocations and a stair-rod
which can be analyzed in terms of standard dislocation theory.
Dissociated loops glide across compact planes and this provides 
a novel mechanism for matter transport away from nanoindentations.
The observation of these defects in other systems, such as
Ag(001), suggests that our results can be generalized to other
(001) fcc metal surfaces, thus providing a consistent description 
with an unprecedented resolution of the incipient plastic 
deformation mechanisms. This might help the interpretation of
elementary yielding events observed in previous nanoindentation 
investigations\cite{nanoAu,nanoAu2,kiely98}.

This research was supported by Spanish CICyT through Project
No. PB96-0652 and by the Office of Basic Energy Sciences, Division of
Materials Sciences, U.S. Department of Energy under contract No.\
DE-AC04-94AL85000. O.R.d.l.F gratefully acknowledges
support from the Spanish MEC.

\begin{figure}
\caption{
(98$\times$98 nm$^2$)
STM image of two nanoindentations in the Au(001) surface.
Rows of {\it hillocks} stemming from the nanoindentation points
and following a \textless110\textgreater\ direction are visible. Bumps of 
pile-up material surround nanoindentation points (the contrast in these
bumps is saturated to enhance the visibility of the {\it hillocks}). Capital 
letters on one of the {\it hillocks} are used to compare their orientations with 
the ones in Fig.~\ref{fig2}.}
\label{fig1}
\end{figure}

\begin{figure}
\caption{a) (25$\times$25 nm$^2$) STM image in Au(001) of a {\it hillock},
such as seen near nanoindentation points. 
b) Scheme of the dislocation configuration proposed for dissociated loops.
Burgers vectors in the Thompson tetrahedron notation and line
directions are shown for each segment.
c) (11.6$\times$11.6 nm$^2$) A {\it hillock} in a Ag(001) surface previously 
ion-bombarded and annealed. Note the atomic resolution and the positions of the 
emerging partial dislocations.
}
\label{fig2}
\end{figure}

\begin{figure}
\caption{Top view of the nanoindentation simulation. 
The dark region in the middle
of the picture corresponds to the indentation point.
Two {\it hillocks} are emitted along \textless110\textgreater\ 
directions. Note the striking similarity of these simulated 
defects with the experimental image of Fig.~\ref{fig2}c.}
\label{fig3}
\end{figure}

\begin{figure}
\caption{(color) 
3D simulation side views from different orientations of the 
dissociated dislocation loop (corresponding to the sub-surface configuration 
of Fig.~\ref{fig3}), colored according to the slip vector. Atoms which signify a 
stacking-fault plane are colored green. Blue and yellow atoms define the core 
of the leading and the trailing partial dislocations, respectively. Red designs
atoms which have slipped a full \textless110\textgreater\ vector. Black arrows 
indicate the indentation point and its axis. The bottom picture can be obtained 
rotating the top one about $90^{\circ}$ counterclockwise around the nanoindentation 
axis.
}
\label{fig4}
\end{figure}


\begin{references}
\bibitem{thompson}
In terms of the Thompson's tetrahedron notation,
the perfect loop segment on the $\overline{a}$
plane would split in this plane under the reaction
\begin{math}
\rm \mbox{{\boldmath $CD$}} \rightarrow \mbox{{\boldmath $C\alpha$}} + 
\mbox{{\boldmath$\alpha D$}}
\end{math}.
The dissociation of both segments on the $\overline{a}$
and $\overline{b}$ planes, respectively, gives rise to the
stacking faults intersecting at the stair-rod dislocation
with Burgers vector {\boldmath$\alpha\beta$}.
\bibitem{details}
 The cell dimensions are 20.4 nm $\times$
20.4 nm $\times$ 12.5 nm with 300.000 atoms. The top
surface is traction free, while the bottom is held fixed with
periodic boundary conditions applied to the side surfaces.
\end{references}
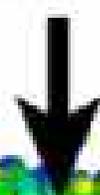
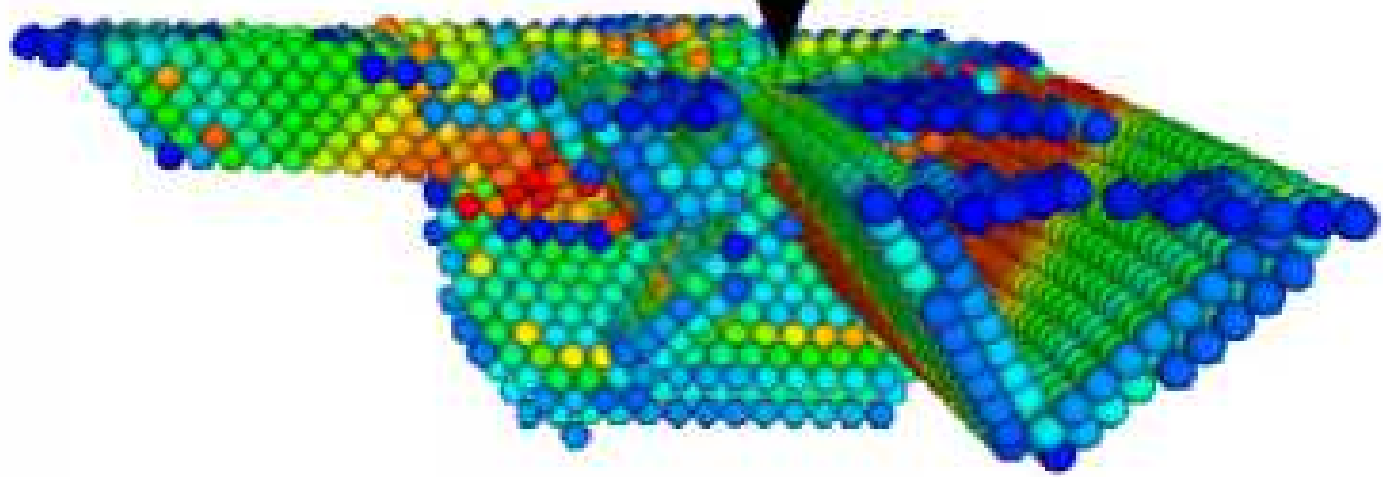
\end{document}